\documentclass[twocolumn]{aastex631}

\usepackage{amsmath}
\usepackage{graphicx}
\usepackage{float}
\usepackage{times}

\begin{document}
	
\title{Physics-Informed Priors Improve Gravitational-Wave Constraints on
  Neutron-Star Matter}
%% \title{Physics-Informed Priors Improve Gravitational-Wave Constraints on
%%   the Nuclear Equation of State \\
%% %
%%   Importance of Physics-Informed Priors to Improve Constraints on
%%   Neutron-Star Matter and Source Classification}

\author[0009-0000-7037-1809]{Spencer J. Magnall}
\affiliation{School of Physics and Astronomy, Monash University, Clayton
  VIC 3800, Australia}
\affiliation{OzGrav: The ARC Centre of Excellence for Gravitational Wave
  Discovery, Clayton VIC 3800, Australia}

\author[0000-0002-8669-4300]{Christian Ecker}
\affiliation{Institut f\"ur Theoretische Physik, Max-von-Laue-Strasse 1,
  60438 Frankfurt, Germany}

\author[0000-0002-1330-7103]{Luciano Rezzolla}
\affiliation{Institut f\"ur Theoretische Physik, Max-von-Laue-Strasse 1,
    60438 Frankfurt, Germany}
\affiliation{School of Mathematics, Trinity College, Dublin 2, Ireland}
\affiliation{CERN, Theoretical Physics Department, 1211 Geneva 23,
  Switzerland}

\author[0000-0003-3763-1386]{Paul D. Lasky}
\affiliation{School of Physics and Astronomy, Monash University, Clayton
  VIC 3800, Australia}
\affiliation{OzGrav: The ARC Centre of Excellence for Gravitational Wave Discovery, Clayton VIC 3800, Australia}

\author[0000-0002-9575-5152]{Simon R. Goode}
\affiliation{School of Physics and Astronomy, Monash University, Clayton
  VIC 3800, Australia}
\affiliation{OzGrav: The ARC Centre of Excellence for Gravitational Wave Discovery, Clayton VIC 3800, Australia}

\begin{abstract}
 Gravitational-wave astronomy shows great promise in determining nuclear
 physics in a regime not accessible to terrestrial experiments. We
 introduce physics-informed priors constrained by nuclear theory and
 perturbative Quantum Chromodynamics calculations, as well as
 astrophysical measurements of neutron-star masses and radii. When these
 priors are used in gravitational-wave astrophysical inference, we show a
 significant improvement on nuclear equation of state constraints.
 Applying these to the first observed gravitational-wave binary
 neutron-star merger GW170817, the constraints on the radius of a
 $1.4\,M_\odot$ neutron star improve from $R_{1.4} =
 {12.54^{+1.05}_{-1.54}} \, {\rm km}$ to $R_{1.4} = 12.11^{+0.91}_{-1.11}
 \,{\rm km}$ and those on the tidal deformability from
 $\tilde{\Lambda}_{1.186} < 720$ to $\tilde{\Lambda}_{1.186} =
 384^{+306}_{-158}$ ($90\%$ confidence intervals) at the events measured
 chirp mass $\mathcal{M}=1.186\,M_\odot$. We also show these priors can
 be used to perform model selection between binary neutron star and
 neutron star-black hole mergers; in the case of GW190425, the results
 provide only marginal evidence with a Bayes factor $\mathcal{BF}=1.33$
 in favour of the binary neutron star merger hypothesis. Given their
 ability to improve the astrophysical inference of binary mergers involving
 neutron stars, we advocate for these physics-informed priors to be used
 as standard in the literature and provide open-source code for
 reproducibility and adaptation of the method.
\end{abstract}
	
		
%========================================================================
\section{Introduction}
\label{sec:Intro}
%========================================================================
	
One of the main goals of gravitational wave (GW) astronomy is to infer
constraints on the neutron-star equation of state (EOS) from observations
of binary neutron star (BNS) mergers. The baryon density in neutron stars
and their mergers can reach several times the nuclear saturation density
$n_s=0.16\,{\rm fm}^{-3}$ and temperatures of the order of tens of
MeV~\citep[see, e.g.,][for reviews]{Baiotti2016,
  Paschalidis2016,Radice2020b}. Such densities and temperatures are
inaccessible to terrestrial experiments and also to \textit{ab initio}
calculations due to the infamous fermionic sign problem, which prevents
direct Quantum Chromodynamics (QCD) solutions on the lattice.

A common approach to astrophysical GW inference is to use agnostic priors
for some GW observables. For example, analyses of the first two observed
BNS mergers GW170817 \citep[e.g.,][]{abbott17_170817observation,
  abbott18_170817EOS, abbott19_170817Properties} and GW190425
\citep[e.g.,][]{abbott20_190425} employed uniform priors on the chirp
mass $\mathcal{M}$ and the dimensionless tidal deformability
$\tilde{\Lambda}$, implicitly assuming these parameters are
uncorrelated. These are the two most informative parameters for the
neutron star EOS.

The above prior assumptions are overly conservative. Basic physical
principles such as causality, consistency with nuclear-physics
experiments, and perturbative QCD calculations, imply a strong
correlation between $\mathcal{M}$ and $\tilde{\Lambda}$ in neutron-star
binaries. Constraints from astrophysical measurements of neutron-star
masses and radii further tighten this correlation~\citep{Altiparmak:2022,
  Ecker:2022, Ecker:2022dlg}.

Motivated by this, we introduce physics-informed joint priors on
$\mathcal{M}$ and $\tilde{\Lambda}$ that incorporate microphysical
constraints from nuclear theory and perturbative QCD, as well as
astrophysical constraints on neutron-star masses and radii into
astrophysical GW parameter inference. We show a significant improvement
on the measurement of the marginalised posterior on $\tilde{\Lambda}$
from the GW observation of GW170817 from $\tilde{\Lambda}_{1.186} < 720$
with the uniform priors to $\tilde{\Lambda}_{1.186} = 384^{+306}_{-158}$
with the new physics-informed prior (confidence intervals are $90\%$
throughout).

We infer a posterior distribution on EOS parameters and find that the
neutron-star radius measurement for a $1.4\,{M_\odot}$ star improves from
${R_{1.4}}=12.54^{+1.05}_{-1.54}\,{\rm km}$, as reported in the original analysis~\citep{abbott19_170817Properties}, to
${R_{1.4}}=12.11^{+0.91}_{-1.11}\,{\rm km}$ when using physics-informed
priors and a spectral EOS~\citep[see also][for some of the first
  estimates made after the GW170817
  event]{Most2018,Capano2020}. Similarly, for GW190425, the effective
tidal deformability improves from $\tilde{\Lambda}_{1.44} =
381^{+1065}_{-271}$ to $\tilde{\Lambda}_{1.44} = 183^{+180}_{-100}$.

In addition to astrophysical inference, we show that these priors can be
used to perform model selection, and hence distinguish between GW-source
types. We perform model selection between a BNS and a neutron star --
black hole (NSBH) origin for GW190425 and find a Bayes factor of $1.33$
in favour of a BNS origin. When including observed rates as the prior odds,
we find an odds ratio of $0.33$, when considering the full population of
NSBH mergers, indicating marginal support for a NSBH origin.  When
considering only mass-gap NSBH mergers, the odds ratio is $1.33$,
marginally in favour of a BNS merger origin. Regardless, we find that
model selection for the origin of GW190425 remains uninformative.
	
In advocating for this prior becoming standard in the literature, we
provide an open-source configuration for the \texttt{Bilby} Bayesian
inference library~\citep{ashton19,romeroshaw20} currently used for the
majority of GW parameter estimation by the LIGO-Virgo-KAGRA (LVK)
collaboration~\citep[e.g.,][]{LIGO, Virgo, KAGRA}. This code provides a
numerical implementation of the joint probability distribution as a
\texttt{Bilby} constrained-prior object for $\mathcal{M}$ and
$\tilde{\Lambda}$. We describe details of the method and implementation
in the next Section.

%========================================================================
\section{Methods}
\label{sec:Methods}
%========================================================================
	
The construction of our physics-informed prior closely follows that
of~\citet{Altiparmak:2022}, which we briefly review here~\citep[see][for
  additional details]{Ecker:2022, Ecker:2022dlg}. We begin by building a
large set of EOSs by stitching together various components. At the
lowest densities ($n<0.5\,n_s$) we adopt the Baym-Pethick-Sutherland
(BPS) model~\citep{Baym71} for the neutron-star crust. In the range
$0.5\,n_s\le n < 1.1\,n_s$, we randomly sample polytropes to span the
range between the softest and stiffest EOSs
from~\citet{Hebeler:2013nza}. At high densities ($n \gtrsim 40\,n_s$),
corresponding to a baryon chemical potential of $\mu = 2.6\,\rm GeV$, we
impose the perturbative QCD constraint from~\citet{Fraga2014} on the
pressure $p(X, \mu)$ of cold quark matter, with the renormalization scale
parameter $X$ sampled uniformly in the range $[1,4]$.
	
For the intermediate density range ($1.1\,n_s < n \lesssim 40\,n_s$), we
use the parametrization method of~\citet{Annala2019}, which models the
sound speed as a function of the chemical potential, $c_s^2(\mu)$, using
piecewise-linear segments:
\begin{equation} \label{eq:cs2}
 c_s^2(\mu) = \frac{\left(\mu_{i+1}-\mu \right) c_{s,i}^2 + \left(\mu -
   \mu_i \right) c_{s,i+1}^2}{\mu_{i+1}-\mu_i}\,,
\end{equation}
where $\mu_i$ and $c_{s,i}^2$ are parameters defining the $i$-th segment
in the range $\mu_i \leq \mu \leq \mu_{i+1}$. In this section, we
adopt natural units where $c=G=1$, unless otherwise stated.

The number density follows as 
\begin{equation} \label{eq:n}
 n(\mu) = n_1 \exp \left({\int_{\mu_1}^\mu \frac{d\mu^\prime}{\mu^\prime
     c_s^2(\mu^\prime)}}\right)\,,
\end{equation} 
where $n_1 = 1.1\,n_s$, and $\mu_1 = \mu(n_1)$ is set by the
corresponding polytropic EOS. The pressure is then obtained via
\begin{equation} \label{eq:p}
 p(\mu) = p_1 + \int_{\mu_1}^\mu d\mu^\prime \, n(\mu^\prime)\,,
\end{equation}
where the integration constant $p_1$ matches the pressure of the
polytrope at $n = n_1$. We numerically integrate Eq.~\eqref{eq:p} using
seven segments for $c_s^2(\mu)$.

Using this framework, we generate approximately $10^6$ EOSs by randomly
selecting the maximum sound speed $c_{s,{\rm max}}^2 \in [0,1]$ and
uniformly sampling the free parameters $\mu_i \in [\mu_1, \mu_{N+1}]$
(where $\mu_{N+1} = 2.6\,\rm GeV$) and $c_{s,i}^2 \in [0, c_{s,{\rm
      max}}^2]$. These EOSs are, by construction, consistent with
nuclear theory and perturbative QCD uncertainties.

To incorporate astrophysical constraints, we solve the
Tolman-Oppenheimer-Volkoff (TOV) equations for each EOS and retain only
those satisfying the mass measurements of J0348+0432~\citep[][$M =
  2.01\pm 0.04\,M_{\odot}$]{Antoniadis2013} and J0740+6620~\citep[][$M =
  2.08 \pm 0.07\,M_{\odot}$]{Cromartie2019, Fonseca2021}, discarding EOSs
with a maximum mass $M_{\rm TOV} < 2.0\,M_{\odot}$. In addition, we
impose NICER radius constraints from J0740+6620~\citep{Miller2021,
  Riley2021} and J0030+0451~\citep{Riley2019, MCMiller2019b}, rejecting
EOSs with $R < 10.75\,\rm km$ at $M = 2.0\,M_{\odot}$ or $R < 10.8\,\rm
km$ at $M = 1.1\,M_{\odot}$.
Naively, we would expect a loss of information since we are discarding the uncertainties associated with these measurements. 
However, \citet{2023Jiang} showed that the underlying distirbutions are almost identical regardless of the choice of constant or variable likelihood,
with 90\% credible intervals essentially overlapping. 
We note that, unlike what was done
in~\citet{Altiparmak:2022} and \citet{Ecker:2022dlg}, we do not impose any
GW-informed constraints on the binary tidal deformability in the
construction of our prior as we are reanalysing these data using the
new prior.
	
In the BNS case, the binary tidal deformability of the prior is computed
as
\begin{equation} 
 \tilde{\Lambda}_{\rm BNS} := \frac{16}{13} \frac{ (12M_2 + M_1) M_1^4
   \Lambda_1 + (12M_1 + M_2) M_2^4 \Lambda_2}{(M_1 + M_2)^5}\,,
\end{equation}
where $M_i$, $R_i$, and $\Lambda_i := \frac{2}{3} k_2 \left( R_i/M_i
\right)^5$ are the component source-frame masses, radii, and tidal
deformabilities, respectively, and $k_2$ is the second tidal Love
number. The source-frame chirp mass is $\mathcal{M} := (M_1
M_2)^{3/5} (M_1+M_2)^{-1/5}$ and we define the mass ratio $q := M_2/M_1$
with $M_2\le M_1$.
	
The tidal deformability of a black hole is zero $\Lambda_{\rm BH}=0$,
implying the total tidal deformability for an NSBH binary is
\begin{equation} 
 \tilde{\Lambda}_{{\rm NSBH}} := \frac{16}{13} \frac{ (12M_{1} + M_{2})
   M_{2}^4 \Lambda_{2}}{(M_{2} + M_{1})^5}\,,
 \end{equation}
where $M_{2}$ and $\Lambda_{2}$ are the neutron-star mass and
tidal deformability, respectively, and $M_{1}$ is the black-hole
mass. 

%------------------------------------------------------------------------
\subsection{BNS and NSBH mergers}
%------------------------------------------------------------------------

\begin{figure}
  \centering
  \includegraphics[width=1.\columnwidth]{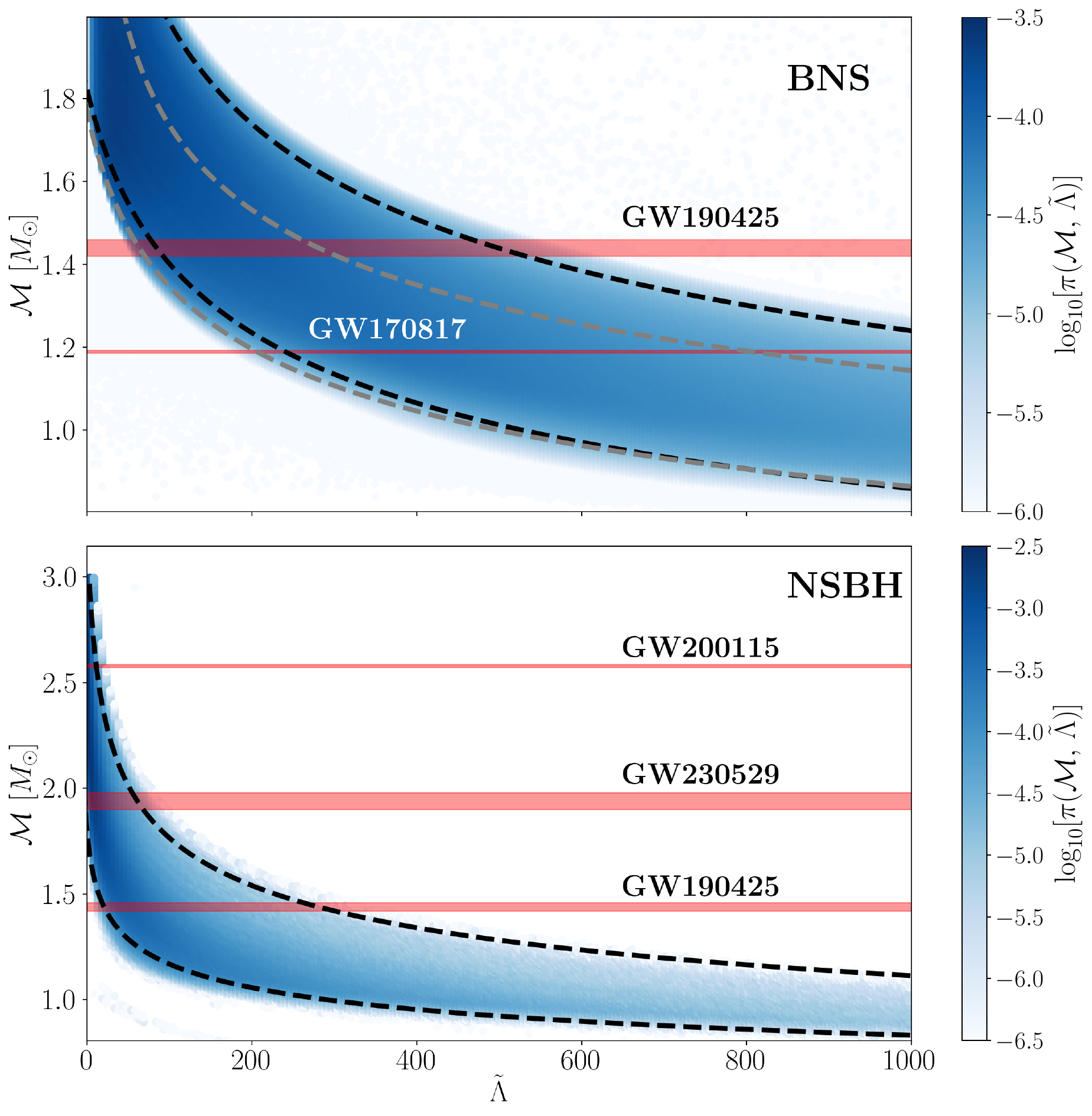}
  \caption{Priors for the source-frame chirp mass $\mathcal{M}$ and the
    dimensionless tidal deformability $\tilde{\Lambda}$ in the case of
    BNS (top panel) or NSBH binaries (bottom panel). In both panels,
    dashed black curves indicate analytic fits to the $90\%$ contours of
    the distribution given in Eq.~\eqref{eq:fit}, while the dashed grey
    curves are the corresponding fits obtained when imposing the GW170817
    constraint on the prior
    following~\cite{Altiparmak:2022}. Horizontal red bands represent the
    $90\%$ credible intervals for the measured chirp masses of
    GW170817~\citep{abbott17_170817observation},
    GW190425~\citep{abbott20_190425}, GW200115~\citep{GW200115}, and
    GW230529~\citep{GW230529}.}
  \label{fig:priors}
\end{figure}

We use the probability distributions described in the previous Section as
priors on the source-frame chirp mass and effective tidal deformability.
These priors are shown in Fig.~\ref{fig:priors} where the top panel is
the BNS prior and the bottom panel is the NSBH prior.  We implement this
in \texttt{Bilby} using the \texttt{conditional\_prior} function,
interpolating a $200\times200$ grid to draw from a joint prior on
$\mathcal{M}$ and $\tilde{\Lambda}$. The $90\%$ probability contours
(dashed black curves) can be fit by a simple power law
\begin{equation}\label{eq:fit}
  \tilde\Lambda_{{\rm min}(\rm max)}=a + b \,\mathcal{M}^c\,,
\end{equation}
where the fitting coefficients are $a=-57\,(-23)$, $b=583\,(2690)$ and
$c=-3.9\,(-4.5)$ for the BNS case and $a=6\,(-5)$, $b=296\,(1700)$ and
$c=-6.6\,(-4.9)$ for the NSBH case. For comparison we also show contours
for $a=-41\,(-0.55)$, $b=539\,(2091)$ and $c=-4.5\,(-5.5)$ (dashed grey
curves) corresponding to a prior in which the constraint on the tidal
deformability from the GW170817 event, i.e., $\tilde{\Lambda} < 720$, is
imposed; this corresponds to the analysis in~\citet{Altiparmak:2022} and
highlights that while the estimate of $\tilde\Lambda_{\rm min}$ is hardly
affected, that for $\tilde\Lambda_{\rm max}$ is significantly smaller
when imposing the GW170817 constraint.

We sample directly in source-frame chirp mass, mass ratio and the two
tidal deformability parameters $\tilde{\Lambda}$ and
$\delta\Lambda$.\footnote{The quantity $\delta\tilde{\Lambda}$ is another
tidal deformability parameter that enters at sixth-order post-Newtonian,
cf. $\tilde{\Lambda}$ that enters at fifth order~\citep[see][for
  definitions of the these quantities]{favata14,wade14}.}  We note that
$q$ is not a variable that depends on the EOS, and we therefore take the
standard approach of choosing a prior on $q$ uniform in the range
$q\in[0.125,\,1]$~\citep[e.g.,][]{romeroshaw20}. The choice of a lower
limit of $q=0.125$ is due to waveform limitations and is not
astrophysically motivated.

While both $\tilde{\Lambda}$ and $\delta\tilde{\Lambda}$ are
related to the EOS, we only impose a prior here on $\tilde{\Lambda}$, as
$\tilde{\Lambda}$ is measured much better than
$\delta\tilde{\Lambda}$~\citep[e.g.,][]{wade14}. In principle, one could
set a combined prior on the three parameters $(\mathcal{M},\,
\tilde{\Lambda}, \,\delta\tilde{\Lambda})$, however, adding this latter
parameter would not add significantly to parameter estimation in the
relatively low signal-to-noise regime. We therefore leave exploration of
physically-motivated priors on $\delta\tilde{\Lambda}$ for future work.

We construct priors on $\mathcal{M}$ and $\tilde{\Lambda}$ by randomly
generating neutron star binaries. This is equivalent to choosing flat
priors on the mass distribution of BNSs. It has been shown that an incorrect choice of neutron star population leads to a bias in the inferred EOS from
gravitational-wave measurements after around 20 observations~\citep[e.g.,][]{2015Agathos,2020Wysocki,2020Landry}. However, given the small number of observed
events we expect any bias to be minimal. We plan to explore
the impact of different mass populations in future work.

The above discussion only refers to BNS mergers as the corresponding
parameterisation does not go to large enough chirp-mass values for
NSBHs. Hence, in the case of NSBH mergers we derive a new distribution by
going back to the raw distributions of progenitor stellar masses and
radii, and combine that progenitor distribution with that of a BH of mass
$M_1$ and tidal deformability $\Lambda_1=0$. In principle, in this case
we could use more sophisticated and realistic mass-ratio distributions,
such as those informed by state-of-the-art binary population synthesis
\citep[e.g.,][]{2021Broekgaarden+}. In practice, we use a
uniformly-distributed mass-ratio prior as we find that our results do not
depend significantly on the choice made for the mass-ratio prior
distribution. Finally, also for NSBHs we choose to work with priors in
$\mathcal{M}\!- \!\tilde{\Lambda}$, enforcing
$\tilde{\Lambda}(\Lambda_2,\mathcal{M},q)$ where $\Lambda_2$ is the tidal
deformability of the neutron star.

%------------------------------------------------------------------------
\subsection{Parameter estimation}
%------------------------------------------------------------------------

\begin{figure}
  \centering
  \includegraphics[width=1.\columnwidth]{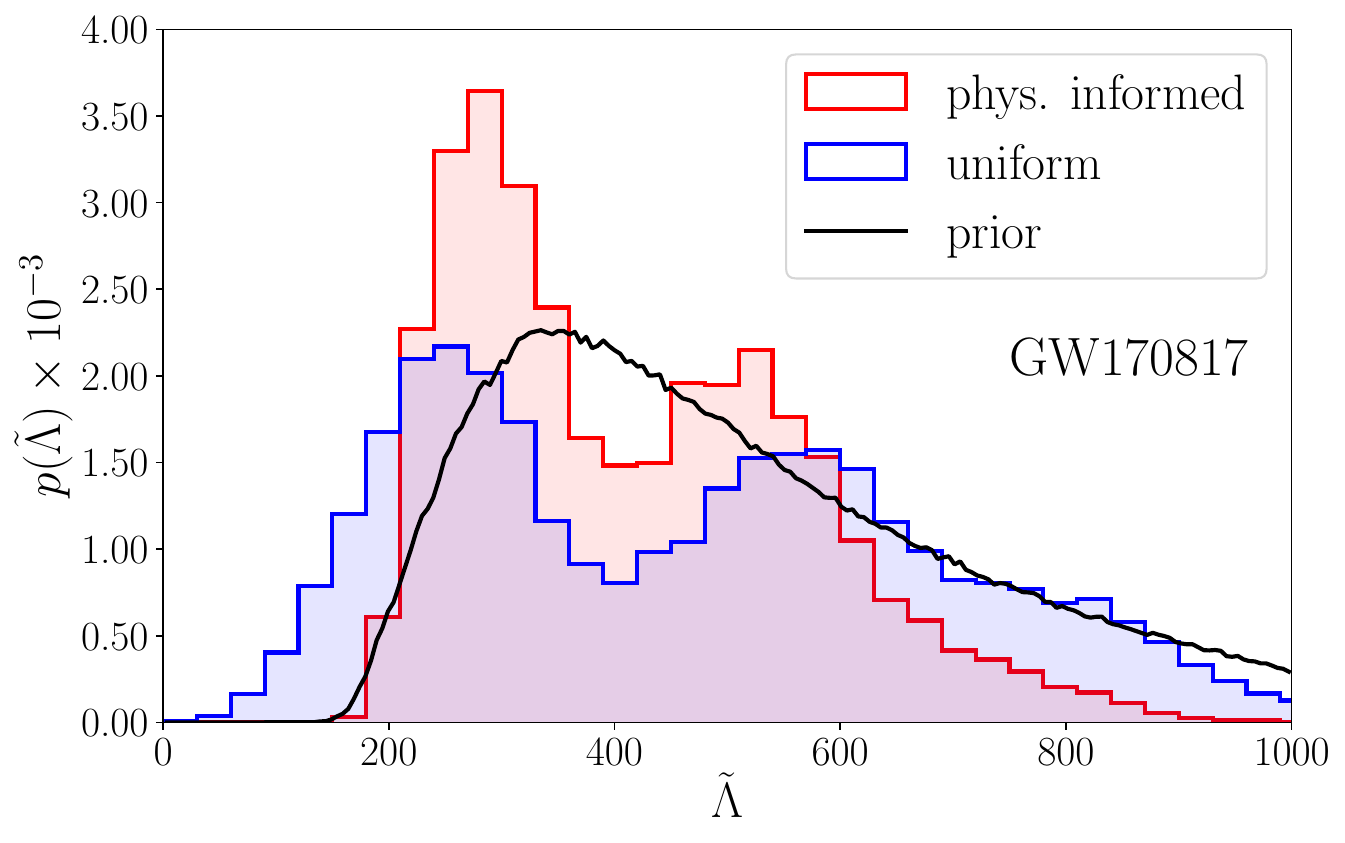}
  \caption{Marginalised posterior distributions of the tidal
    deformability $\tilde{\Lambda}$ for GW170817. The posterior using a
    physics-informed prior is shown in red, while the posterior using
    uniform priors from the LVK analysis is shown in blue: we show the
    (marginalised) prior with the black curve. Note that the posterior
    from the LVK shows support up to $\tilde{\Lambda} = 2000$.}
  \label{fig:GW170817_tides}
\end{figure}	

We perform Bayesian parameter estimation on the GW strain data of the two
BNS merger events GW170817 and GW190425 using the publicly-available data
from the GWOSC repository\footnote{\url{https://gwosc.org}}. We utilise
the \texttt{Bilby} Bayesian inference
framework~\citep{ashton19,romeroshaw20} and priors on source--frame chirp
mass and tidal--deformability as discussed in Section
\ref{sec:Methods}. Priors for all remaining parameters assume the
so-called low-spin ($\chi \leq 0.05$) default priors as defined in
\texttt{Bilby}. We use Planck 2018 cosmological parameters~\citep{Planck}
such that Hubble's constant and the matter density are respectively
$H_0=67.66 \rm \ km \ s^{-1} \ Mpc^{-1}$, $\Omega_M = 0.30966$.
	
Parameter estimation for each run is performed using the \texttt{dynesty}
nested sampler~\citep{Dynesty} and the \texttt{IMRPhenomPv2\_NRTidal}
waveform~\citep{IMRPhenomP_NRtidal}. Analysis of BNS merger signals is
computationally expensive and as such we utilise parallel implementation
of \texttt{Bilby}, i.e., \texttt{pBilby}~\citep{pbilby}. We reconstruct
pressure--density and mass--radius curves from posterior samples of
source--frame chirp mass, mass ratio, and effective tidal deformability
by performing EOS inference on the posterior samples of mass and tidal
deformability. We utilise a four-piece spectral
parameterisation~\citep{Lindblom2010} stitched to an
SLy~\citep{2001_SLy_EOS} crust at low densities (further details can be
found in Appendix~\ref{sec:EOS_inference}).

The choice of adopting an approach for the EOS inference (i.e.,
four-piece spectral parameterisation) that is different from that used in
the EOS construction (i.e., seven-segment sound-speed parameterisation)
follows a precise logic. In particular, this has the advantage that it is
possible to compare directly with the LVK inference analysis and hence
clearly distinguish the assumptions behind the definition of the priors
from those employed in the inference. At the same time, this also implies
that the posteriors obtained for the EOSs parameters differ from the
corresponding priors. While these differences are small, we plan to
explore the use of the same set of assumptions for the EOS priors and
inference in future work.

%========================================================================
\section{Results}
\label{sec:Results}
%========================================================================
	
\begin{figure}
  \centering
  \includegraphics[width=1.\columnwidth]{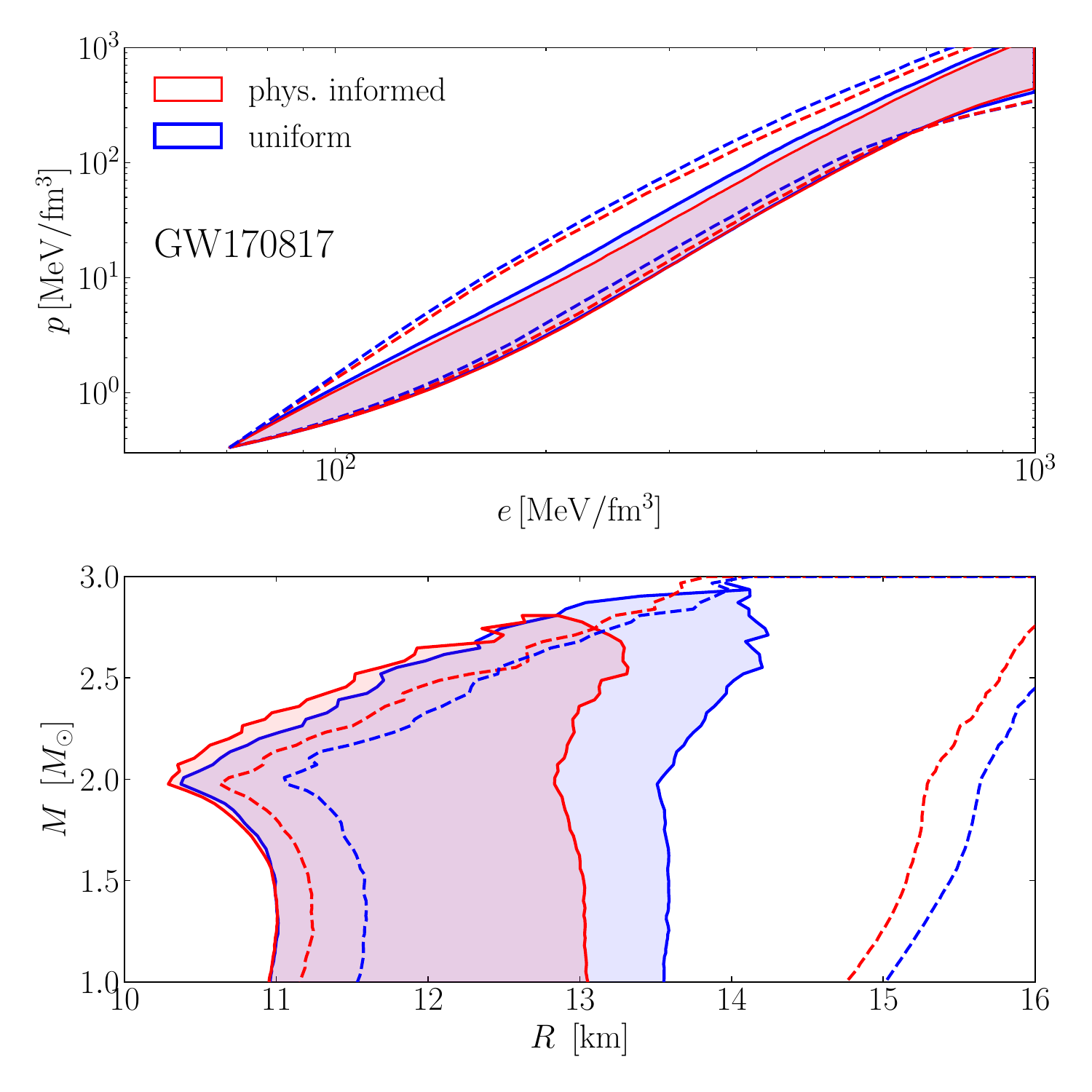}
  \caption{Posterior distributions of the range of pressures and energy
    densities for a nuclear-matter EOS (top panel) and of the
    corresponding masses and radii (bottom panel) in the case of
    GW170817. In both panels, the shaded regions show the $90\%$ credible
    intervals of the posterior distributions obtained using the
    physics-informed priors (red) or the uniform priors (blue), while the
    dashed curves show the $90\%$ credible intervals of the priors on
    chirp mass and tidal deformability when mapped to the spectral
    parameterisation.}
    \label{fig:GW170817_massradius}
\end{figure}

Figure \ref{fig:GW170817_tides} shows the posterior distributions of the
tidal deformability of GW170817 analyzed using the physics-informed BNS
(red), the publicly-released uniform priors from the LVK
collaboration~\citep[blue;][]{abbott17_170817observation}, and the
marginalised prior for $\tilde{\Lambda}$ (black). While we show only the
tidal posteriors, we sample over all binary parameters. Posterior
distributions for all other binary parameters, including the chirp mass
and mass ratio, are identical for the physics-informed and uniform-prior
runs.  Note that, as expected, the marginalised posterior is not
significantly different at large tidal deformabilities (compare blue and
black curves), but shows significant changes at small values of
$\tilde{\Lambda}$, essentially excluding values $\tilde{\Lambda} \lesssim
200$ for the physics informed priors. As a result, the $90\%$ lower and
upper limits on the tidal deformability are now $226 \leq
\tilde{\Lambda}_{1.186} \leq 690$. These arguably represent the most
interesting consequence of the physics-informed prior on the
$\mathcal{M}\!-\!\tilde{\Lambda}$ relation.

Figure~\ref{fig:GW170817_massradius} shows the impact of the
$\mathcal{M}\!-\!\tilde{\Lambda}$ relation on the reconstructed posterior
distributions of the pressure and energy density (top panel) and of the
mass and radius (bottom panel) relative to the GW170817 event. More
specifically, the shaded regions show the $90\%$ credible intervals for
the posteriors samples obtained using the physics-informed priors (red)
and uniform priors (blue). In this way, we can improve the constraints on
the radius of a $1.4\,M_\odot$ neutron star from the values provided by
LVK $R_{1.4} = {12.54^{+1.05}_{-1.54}}\,{\rm km}$ to $R_{1.4} =
12.11^{+0.91}_{-1.11}\,{\rm km}$. The dashed curves in
Fig.~\ref{fig:GW170817_massradius} represent the $90\%$ credible
intervals of the priors on the physics-informed prior (red) and on the
uniform prior (blue) when sampled with the spectral parameterisation.

\begin{figure}
  \centering
  \includegraphics[width=1.\columnwidth]{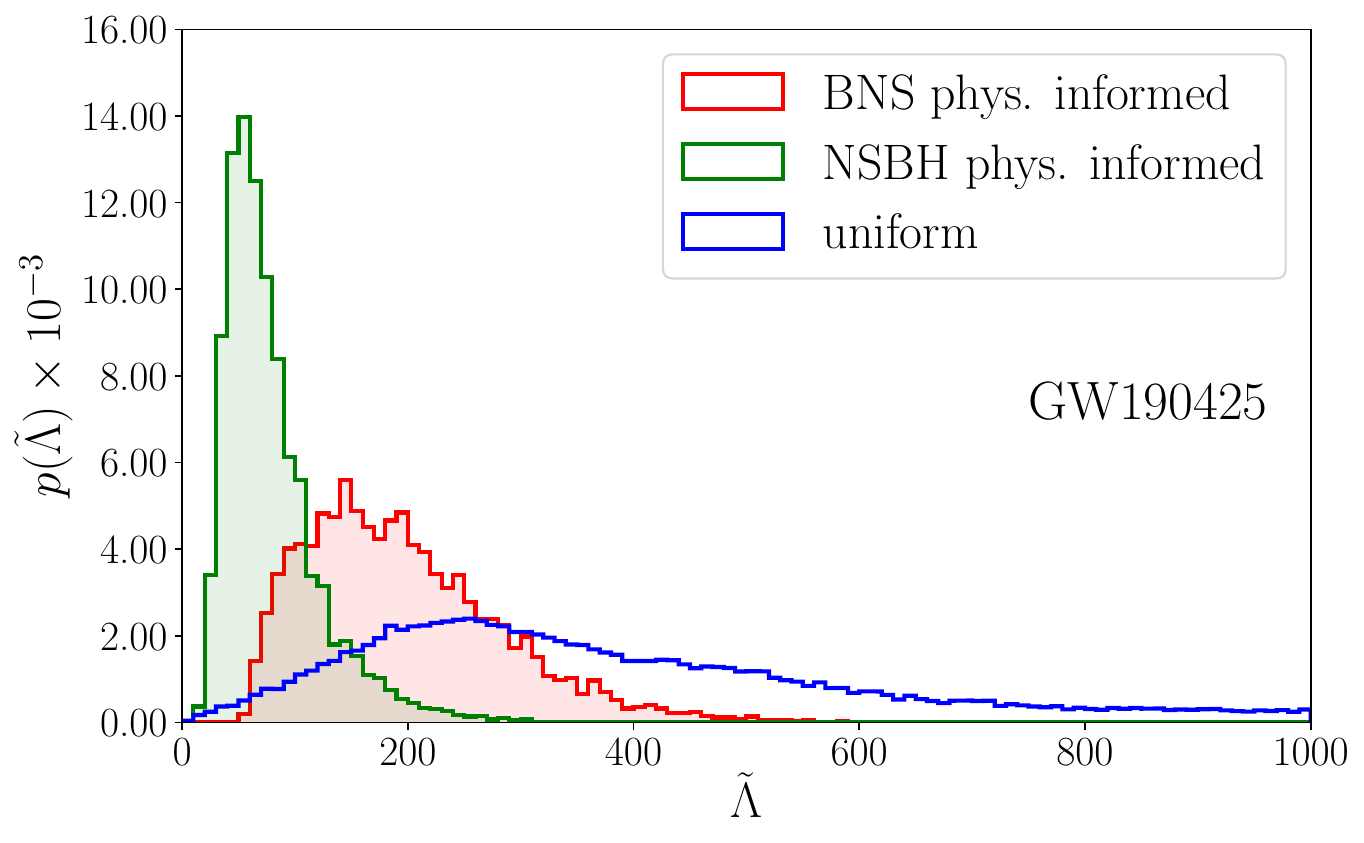}
  \caption{The same as in Fig.~\ref{fig:GW170817_tides} but for the
    GW190425 event. The posterior using a physics-informed prior for BNS
    mergers is shown in red, while the posterior for NSBH mergers is
    shown in green. The posterior using uniform priors from the LVK is
    reported in blue.}
  \label{fig:GW190425_tides}
\end{figure}
\begin{figure}
  \centering
  \includegraphics[width=1.\columnwidth]{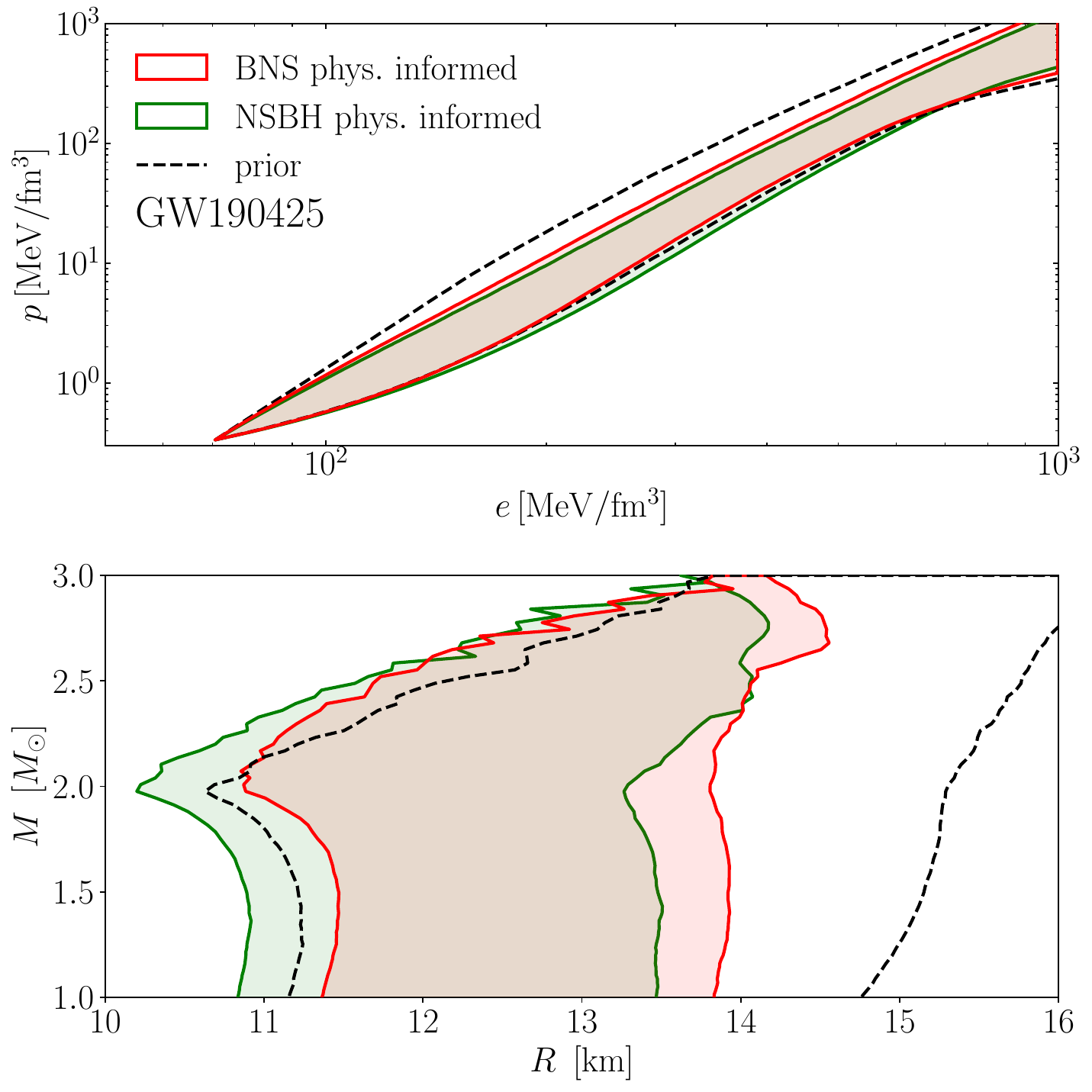}
  \caption{The same as in Fig.~\ref{fig:GW170817_massradius} but for the
    GW190425 event. The top panel shows the posterior distributions in
    pressure--density space, while the bottom panel shows the
    reconstructed mass--radius curves. The shaded regions show the $90\%$
    credible intervals of the posterior distributions obtained from the
    analysis using the physics-informed BNS prior (red) and the NSBH
    prior (green). The black dashed curve shows the $90\%$ credible
    intervals of the priors when
    mapped to a spectral parameterisation.}
  \label{fig:GW190425_massradius_BNS}
\end{figure}
Figure \ref{fig:GW190425_tides} offers a complementary view to
Fig.~\ref{fig:GW170817_tides} and shows the posterior distribution of the
intrinsic tidal parameter of GW190425 when analyzed using both the BNS
prior (red) and NSBH prior (green; see Fig.~\ref{fig:priors}). We show
the posterior distribution of the LVK collaboration obtained
using uniform priors in blue~\citep{abbott20_190425}.

Bayesian-model selection between the BNS and NSBH merger for the GW190425
event gives a Bayes factor of $1.33$.  We calculate the odds ratio, the
product of the Bayes factor and our prior odds ratio. The latter can be
calculated as a ratio of observed events, in our case the ratio of
observed BNS and observed NSBH mergers. The total number of BNS mergers
(excluding GW190425) is one; and the total number of publicly-released
confirmed NSBH mergers is three: GW200105, GW200115~\citep{GW200115},
GW230529~\citep{GW230529}. Thus the prior odds ratio is $1/3$ and the
odds ratio is $0.44$, which represents marginal evidence for a NSBH
merger. To appreciate how marginal this estimate is, we might argue that
only mass-gap events like GW230529 should be included in the prior. In
this case, the prior odds is unity and the odds ratio is again $1.33$, in
favour of a BNS origin. Regardless, any choice of prior odds remains
uninformative for determining the nature of GW190425 and additional
observables [e.g., on the properties of the ejected
  matter~\citep{Most:2021b}] need to be included in the priors.

Figure \ref{fig:GW190425_massradius_BNS} is logically equivalent to
Fig.~\ref{fig:GW170817_massradius}, but contrasting BNS and NSBH merger
interpretations for the GW190425 event. Hence, it shows the reconstructed
pressure--density (top) and mass--radius (bottom) posteriors of
GW190425. The EOS $90\%$ credible intervals of posteriors obtained using
the BNS physics-informed prior are shown in red, while the NSBH
posteriors are shown in green. Once again, the dashed curve represents
our prior on the EOS. Overall, we find that the posterior distributions
constrain $R_{1.4} = 12.69^{+1.24}_{-1.22} \, {\rm km}$ using the BNS
prior and $R_{1.4} = 12.06^{+1.44}_{-1.15} \, {\rm km}$ using the NSBH
prior.

%========================================================================
\section{Conclusions}
%========================================================================

The still poor knowledge of the EOS of nuclear matter represents a
significant obstacle in the analysis of GW events involving neutron
stars. This is true not only for parameter estimation of neutron stars in
the case of BNS mergers, but also to assess whether the low-mass
component is actually a neutron star or a black hole in candidate NSBH
mergers. To counter these limitations, a number of approaches have
recently been developed that use a large number of physically and
astrophysically consistent parameterised EOSs to deduce quasi-universal
behaviours of the neutron star. A particularly useful relation is the one
presented by~\cite{Altiparmak:2022}, who have shown a tight statistical
correlation between the chirp mass $\mathcal{M}$ -- arguably, the
best-measured quantity in binary mergers -- and the binary tidal
deformability $\tilde{\Lambda}$.

We here exploit the $\mathcal{M}\!- \!\tilde{\Lambda}$ relation to
re-investigate the constraints on the GW170817 and GW190425 events. More
specifically, we employ the reference LVK pipeline analysis where uniform
priors are replaced by the physics-informed priors coming from the EOS
parameterisation. In this way, we find that a significant improvement in
the inferred properties is obtained in the case of GW170817, where the
constraints on the radius of a $1.4 \, M_\odot$ neutron star improve from
$R_{1.4} = {12.54^{+1.05}_{-1.54}} \, {\rm km}$ to $R_{1.4} =
12.11^{+0.91}_{-1.11} \, {\rm km}$.  Similarly, the constraints on the
tidal deformability are considerably restricted, going from
$\tilde{\Lambda}_{1.186} < 720$ to $\tilde{\Lambda}_{1.186} =
384^{+306}_{-158}$. In the case of GW190425, we assess the nature of the
secondary in the binary by computing the odds ratio associated to a
neutron star or a black hole hypothesis.  We find this analysis to be
uninformative, with a odds ratio between $0.44$ and $1.33$ in favour of
the BNS merger hypothesis.  In either case, we improve the constraints on
the radius estimates with respect to previous LVK with the posterior
distributions indicating $R_{1.4} = 12.69^{+1.24}_{-1.22} \, {\rm km}$
for BNS and $R_{1.4} = 12.06^{+1.44}_{-1.15} \, {\rm km}$ for NSBH
mergers. We give separate constraints on the EOS for both BNS and NSBH
origins of GW190425; however we could also calculate a single EOS
posterior weighted by the evidence for BNS and NSBH merger origins of
GW190425~(e.g., \citealt{2020Ashton}).
Compared to other works which provide constraints on the cold nuclear EOS by combining nuclear and astrophysical measurements using joint likelihoods
 \citep[e.g.,][]{2019Coughlin,2020Dietrich,2023Tiwari,2023Pang,2024Breschi,2025Koehn,2025Golomb}
we obtain consistent, if slightly broader constraints on the radius of a $1.4 \, M_\odot$ neutron star.

Overall, our results clearly indicate that physics-informed priors do
improve the parameter estimation of GW events involving neutron stars and
we thus advocate their use as versatile tools for which we provide an
open-source code for reproducibility and further adaptation of the
method. At the same time, what we present here can be improved in a
number of ways. First, by adopting for the EOS inference the same EOS
parameterisation employed in the construction of the physics-informed
priors. Second, while we have constructed priors using binaries
containing neutron stars with ``low spins'' (i.e., $\chi:=J/M^2 \leq
0.05$, where $J$ and $M$ are the star's angular momentum and mass,
respectively), the inference can be extended by considering also
``high-spin'' ($\chi \leq 0.89$) priors. Finally, in the case of NSBH
mergers, the posteriors can be further constrained by restricting the
high-range of neutron-star masses that would lead to a prompt black-hole
formation and hence prevent an electromagnetic
counterpart~\citep{Bauswein2013, Koeppel2019, Tootle2021,
  Kashyap:2021wzs, Koelsch2021, Ecker:2024kzs}.  We plan to explore these
extensions in future work.

\appendix

%========================================================================
\section{Equation of state inference}
\label{sec:EOS_inference}
%========================================================================

The posterior distribution on the EOS parameters $\eta$ from the GW data is
\begin{equation}
  p(\epsilon| d) = \frac{p(d|{\epsilon})\pi(\epsilon)}{\mathcal{Z}}\,,
\end{equation}
where $\pi(\epsilon)$ is the prior on the EOS parameters, and
$\mathcal{Z}$ is the evidence. The likelihood is given by
\begin{equation}
    p(d|\epsilon) = \int  d\theta \,p(d|\theta)\,\pi(\theta|\epsilon)\,,
\end{equation}
where $\theta$ are our nuisance parameters which we marginalise over.
Our likelihood reduces to
\begin{equation}
    p(d|\epsilon) = \int \ d\theta \ p(d|\mathcal{M},q,\tilde{\Lambda})
    \ \pi(\mathcal{M},q,\tilde{\Lambda}|\epsilon,\theta) \ d\mathcal{M}
    \ dq \ d \tilde{\Lambda}\,.
\end{equation}
Since we a dealing with a small number of events (i.e., $1-2$), we may
ignore the modelling of overall neutron-star population parameters
without introducing significant biases into the EOS inference (e.g.,
\citealt{2015Agathos, 2020Wysocki, 2020Landry}). The marginalised
likelihood transforms to
\begin{equation}
    p(d|\epsilon) = \int \ p(d|\mathcal{M},q,\tilde{\Lambda})
    \ \pi(\mathcal{M},q,\tilde{\Lambda}|\epsilon)\ d\mathcal{M} \ dq \ d
    \tilde{\Lambda}\,.
    \label{eq:P_d_eps}
\end{equation}

In expression~\eqref{eq:P_d_eps} there are two terms that need to be
calculated. The first one is $\pi(\mathcal{M}, q,
\tilde{\Lambda}|\epsilon)$ that represents a prior on chirp mass,
mass-ratio and $\tilde{\Lambda}$ for a given EOS. The second
one is the likelihood $p(d|\mathcal{M},q,\tilde{\Lambda})$. Typically,
this likelihood would be obtained by reweighing posterior samples of
parameter estimation runs
\begin{equation}
    p(d|\mathcal{M},q,\tilde{\Lambda}) =
    \frac{p(\mathcal{M},q,\tilde{\Lambda}|d)}{\pi_{\rm
        PE}(\mathcal{M},q,\tilde{\Lambda})}\,,
\end{equation}
however if we did this we would be removing any of our prior information
from the posterior samples. Since our prior on $\tilde{\Lambda}$ and
$\mathcal{M}$ is actually physically motivated, we set
\begin{equation}
    \pi_{\rm PE}(\mathcal{M},q,\tilde{\Lambda}) =
    \pi(\mathcal{M},q,\tilde{\Lambda}|\epsilon)\,,
\end{equation}
which then implies that
\begin{equation}
    p(d|\epsilon) = \int p(\mathcal{M},q,\tilde{\Lambda}|d)\ d
    \mathcal{M} \ d q \ d \tilde{\Lambda}\,.
\end{equation}

Since the chirp mass is often very-well constrained by parameter
estimation, we take the median value of chirp mass $\bar{\mathcal{M}}$,
and rewrite our posterior distribution as
\begin{equation}
    p(\mathcal{M},q,\tilde{\Lambda}|d) = p(q,\tilde{\Lambda})
    \ \delta(\mathcal{M}- \bar{\mathcal{M}})\,,
\end{equation}
which simplifies the integral to
\begin{equation}
    p(d|\epsilon) = \int p(\bar{\mathcal{M}},q,\tilde{\Lambda}|d)\ d q
    \ d \tilde{\Lambda}\,.
\end{equation}
We note that we can also express the tidal deformability as a function of
$\bar{\mathcal{M}}$, $q$ and $\epsilon$, which removes another dimension
from our integration, namely,
\begin{equation}
        p(d|\epsilon) = \int
        p(q,\tilde{\Lambda}(\bar{\mathcal{M}},q,\epsilon)|d)\ d q\,,
\end{equation}
which is now just an integral over the mass ratio $q$.

We evaluate the probability density of our marginalised posterior samples
using a bounded kernel-density estimate and parameterise our EOS using
the spectral parameterisation of \cite{Lindblom2010}. We use uniform
priors for each EOS parameter and enforce a minimum TOV mass $M_{\rm TOV}
\geq 2.0 M_{\odot}$ and causality.
	
%============================================================
\section*{Acknowledgements}
%============================================================

The authors are grateful to the organisers of the NEOSGrav2024 (Neutron
Star Equation of State and Gravitational Waves) conference, held in Goa,
India, from October 1-4, 2024, and organised by the Inter-University
Centre for Astronomy and Astrophysics (IUCAA), Pune, where the
discussions that led to this work were initiated. The authors thank Debarati Chatterjee, Nir Guttman,
Floor Broekgaarden and Nikhil Sarin. SJM receives
support from the Australian Government Research Training Program. Partial
funding comes from the ERC Advanced Grant ``JETSET: Launching,
propagation and emission of relativistic jets from binary mergers and
across mass scales'' (Grant No. 884631), and the European Horizon Europe
staff exchange (SE) programme HORIZON-MSCA2021-SE-01 Grant
No. NewFunFiCO-101086251. CE acknowledges support by the Deutsche
Forschungsgemeinschaft (DFG, German Research Foundation) through the
CRC-TR 211 ``Strong-interaction matter under extreme conditions''--
project number 315477589 -- TRR 211. LR acknowledges the Walter Greiner
Gesellschaft zur F\"orderung der physikalischen Grundlagenforschung
e.V. through the Carl W. Fueck Laureatus Chair and the hospitality at
CERN, where part of this research was carried out. SJM, PDL, and SRG are
supported through Australian Research Council (ARC) Centres of Excellence
CE170100004 and CE230900016, Discovery Projects DP220101610 and
DP230103088, and LIEF Project LE210100002.  We also computational support
through the Ngarrgu Tindebeek / OzSTAR Australian national facility at
Swinburne University of Technology. This material is based upon work supported by NSF’s LIGO Laboratory which is a major facility fully funded by the National Science
Foundation.

%------------------------------------------------------------------------
\section*{Data Availability}
%------------------------------------------------------------------------

The implementation of our physics-informed priors into the \texttt{Bilby}
Bayesian inference library, along with documentation and example scripts
can be found at \url{https://github.com/spencermagnall/physics\_informed\_priors}.

\bibliography{references}{}
\bibliographystyle{aasjournal}

\end{document}